\documentclass[9pt]{article}
\usepackage[utf8]{inputenc}
\usepackage{siunitx}
\usepackage[version=3]{mhchem}
\usepackage[a4paper,top=2cm,bottom=3cm,left=2cm,right=2cm,marginparwidth=1.75cm]{geometry}
\usepackage{graphicx}
\usepackage{amsmath}
\usepackage{xfrac}
\usepackage[T1]{fontenc}
\usepackage{authblk}
\usepackage{abstract}

\usepackage{environ}
\NewEnviron{altequation}{%
\begin{equation}
\scalebox{0.91}{$\BODY$}
\end{equation}
}

\title{LED based photoacoustic \ce{NO2} sensor with sub-ppb detection limit}
\author[1]{Juho Karhu}
\author[2]{Tuomas Hieta}
\author[1]{Farshid Manoocheri}
\author[3,4]{Markku Vainio}
\author[1,5]{Erkki Ikonen}

\affil[1]{Metrology Research Institute, Aalto University, Maarintie 8, FI-02150 Espoo, Finland}
\affil[2]{Gasera Ltd., Lemminkäisenkatu 59, FI-20520, Turku, Finland}
\affil[3]{Department of Chemistry, University of Helsinki, P.O. Box 55, FI-00014, Helsinki, Finland}
\affil[4]{Photonics Laboratory, Physics Unit, Tampere University, FI-33014 Tampere, Finland}
\affil[5]{VTT MIKES, VTT Technical Research Centre Finland, P.O. Box 1000, FI-02044 VTT, Finland}
\date{}

\begin{document}
\twocolumn[
\maketitle
\hrule
\begin{abstract}
\noindent A high-sensitivity LED based photoacoustic  \ce{NO2} sensor is demonstrated. Sensitive photoacoustic gas sensors based on incoherent light sources are typically limited by background noise and drifts due to a strong signal generated by light absorbed at the photoacoustic cell walls. Here we reach a sub-ppb detection limit and excellent stability by using cantilever-enhanced photoacoustic detection and performing a two-channel relative measurement. A white-light LED is used as a light source and the spectrum is divided into two wavelength channels with a dichroic filter. The photoacoustic signals generated by the two wavelength channels are measured simultaneously and used to solve the \ce{NO2} concentration. The background signal is highly correlated between the two channels and its variations are suppressed in the relative measurement. A noise level below 1 ppb is reached with an averaging time of 70 s. This is, to the best of our knowledge, the first time a sub-ppb detection limit is demonstrated with an LED based photoacoustic \ce{NO2} sensor. As LEDs are available  at  wide  selection of emission wavelengths, the results show great potential for development of cost-effective and sensitive detectors for a variety of other trace gasses as well.
\end{abstract}
\vspace{4pt}
\hrule
\vspace{10pt}
]

\section{Introduction}
\noindent Nitrogen dioxide (\ce{NO2}) is an atmospheric trace gas with adverse health effects particularly to the respiratory system \cite{EPA2016}. The yearly averages of urban \ce{NO2} concentrations are typically in the order of \SI{10}{ppb}, although near high traffic the local concentrations can be significantly higher \cite{CYRYS12}. Comprehensive \ce{NO2} monitoring requires access to portable and cost-effective sensors, with high sensitivity below ppb-level.

Photoacoustic (PA) detectors have been studied as a candidate for compact and sensitive \ce{NO2} sensors. Sensors based on optical absorption offer excellent sensitivity, selectivity and stability. As opposed to most other sensing techniques based on optical absorption, PA detectors have great potential towards miniaturization, because the PA signal is inversely proportional to the sample volume \cite{Kreuzer77}. Mid-infared laser sources have been used for PA detection of \ce{NO2} \cite{Pushkarsky06,Lamard19}. Detection limit down to \SI{0.5}{ppb} was reached with a quantum cascade laser emitting at the wavelength of the strong vibrational transitions of \ce{NO2} around \SI{6250}{\nano\meter} \cite{Lamard19}, but laser sources at long wavelengths are often expensive and have limited wall-plug efficiencies in the order of a few percent. Fortunately, \ce{NO2} absorbs strongly also in the visible wavelength range, where detection limits below \SI{1}{ppb} have been demonstrated in PA measurements with lasers as light sources  \cite{kalkman08,Pan20,Peltola15,RUCK17,YIN17}. The high sensitivity is achieved using high-power lasers and resonant acoustic cells. PA based \ce{NO2} sensors have also been demonstrated using LED light sources, but the detection limits have been in the range of tens of ppb at best \cite{Saarela11,Kapp19,BERNHARDT10}\footnote{In reference \cite{ZHENG15}, a detection limit of \SI{1.3}{ppb} is reported, but as previously noted in reference \cite{RUCK18}, the light sources appears to have been a laser diode, despite being referred to as an LED in parts of the text}. LED light sources offer good stability, high efficiency, low cost and wide availability of emission wavelengths \cite{pulli15}. The development of LED based PA detectors with detection limits at ppb-level and below is a promising step toward low-cost, efficient and portable optical trace gas sensors for \ce{NO2} and a variety of other compounds.

In sensitive PA detectors with high optical powers, the signal generated from absorption by the PA cell windows or scattered light hitting the cell walls often produces a strong background signal. The background can increase the noise level and degrade the stability of the detector if not properly separated from the concentration signal. This is especially a problem with light from incoherent sources such as LEDs, which is likely to come into contact with the cell walls. In the past LED-based PA measurements, this problem has been addressed by cell designs which attempt to minimize the amount of light hitting the cell walls \cite{Kapp19} and by differential measurements, where another light source is modulated with opposite phase to compensate for the wall signal \cite{BERNHARDT10}.

In this article, we present an LED-based PA detector for \ce{NO2} with detection limit below \SI{1}{ppb}. This is, to the best of our knowledge, the lowest reported detection limit reached with an LED based photoacoustic \ce{NO2} sensor and an order of magnitude improvement to the previous best reports. The high sensitivity in our system is reached by using cantilever-enhanced photoacoustic spectroscopy (CEPAS) \cite{Kuusela07}. We separate the concentration signal and the wall contribution using a simple measurement at two wavelength bands. Our light source is a white-light LED and we divide the spectrum in two with optical filters. Because CEPAS does not require detection at a resonance frequency to reach high sensitivity, we can multiplex the two color channels to two different modulation frequencies. The two signals can thus be detected simultaneously either by a Fourier transform of the CEPAS output or by two phase sensitive measurements at different frequencies. A relative measurement of the two color channels suppresses the problems arising from the strong background signal and leads to improved stability and a better detection limit.

\section{Experimental section}
\begin{figure}[tb]
    \centering
    \includegraphics[width=0.95\linewidth]{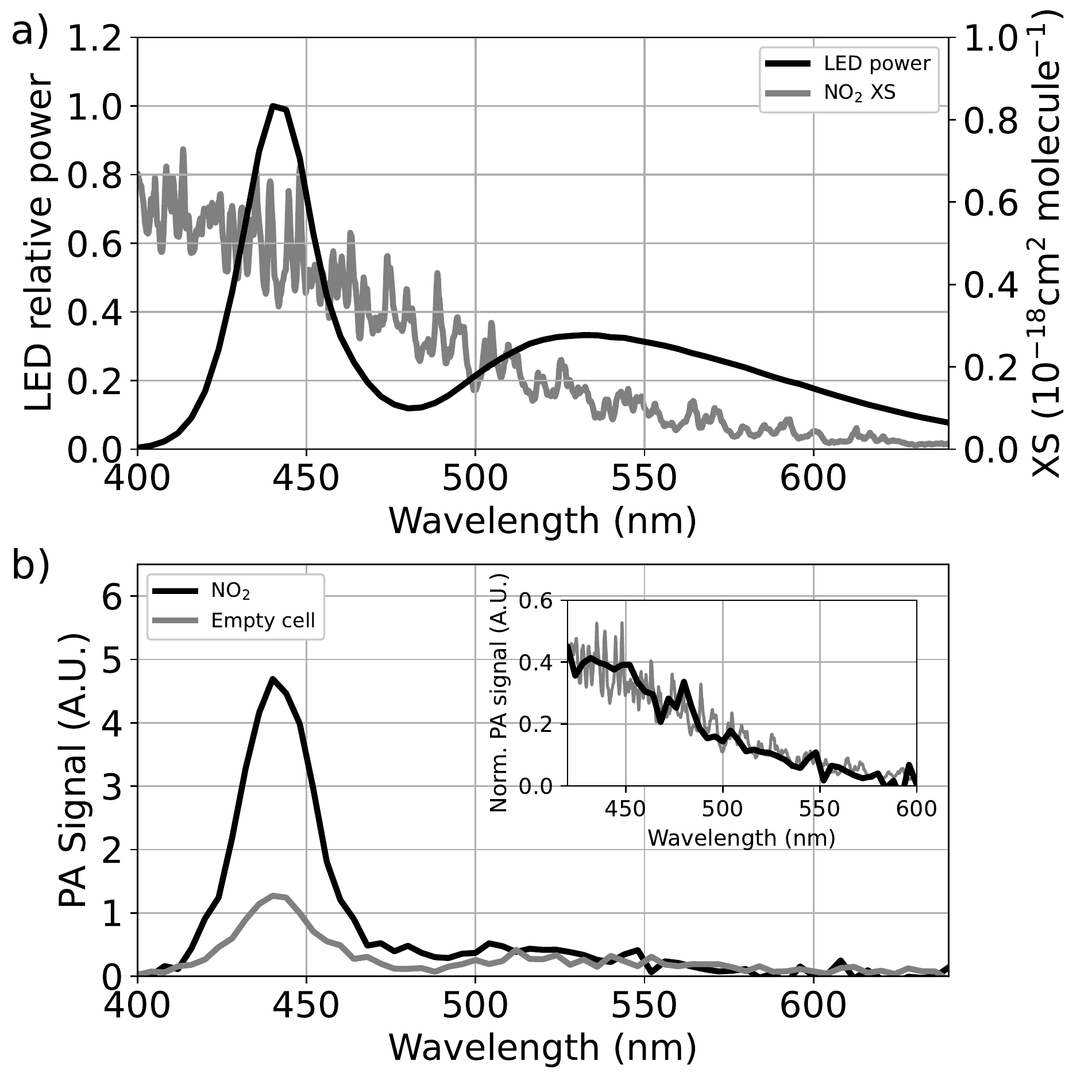}
    \caption{(a) LED emission spectrum measured with the monochromator and \ce{NO2} absorption cross-section (XS) \cite{BURROWS98}. (b) PA spectrum of 10 ppm of \ce{NO2} and an air-filled cell (Empty cell). Inset in (b) shows the power normalized spectrum for \ce{NO2} and the shape of the \ce{NO2} XS for reference.}
    \label{fig:monochrom}
\end{figure}

The CEPAS instrument is similar to Gasera PA201 \cite{Peltola15}, except the cell dimensions are different: the cell length is \SI{40}{\milli\meter} and its diameter is \SI{8}{\milli\meter}. The light source was a commercial white-light LED (MCWHD2, Thorlabs). The LED was fixed to a heat sink, which was kept at \SI{25}{\celsius} with a thermoelectric cooler. Because the LED emission spectrum is wide (figure \ref{fig:monochrom}a) and the photodissociation of \ce{NO2} may affect the relative PA response at shorter wavelengths \cite{Tian13}, we first verified the spectral response with a monochromator measurement. The LED light was collected and focused into the monochromator (TMc300, Bentham Instruments) with a pair of lenses. The LED emission power was amplitude modulated at \SI{90}{\hertz} with a mechanical chopper (SR540, Stanford Research Systems). After the monochromator, the light was collected with a parabolic mirror and focused into the CEPAS cell. The light transmitted through the cell was focused onto a silicon photodetector (PDA100A2, Thorlabs). The CEPAS signal was digitized with a data acquisition card (NI USB-6356, National Instruments) simultaneously with a reference signal from the chopper. The signal processing and measurement automation was performed with a LabVIEW program. Phase-sensitive detection of the CEPAS signal was performed digitally at the measured frequency of the reference signal. The spectrum was measured from a sample of \SI{10}{ppm} of \ce{NO2} in nitrogen at a sample pressure of \SI{600}{\milli\bar}. The sample inside the CEPAS cell was renewed every \SI{25}{\second}.

The photoacoustic response of \ce{NO2} as a function of the monochromator bandpass wavelength is shown in figure \ref{fig:monochrom}b. The monochromator step size and bandwidth were set to \SI{4}{\nano\meter}. Each wavelength was averaged for \SI{5}{\second}. A spectrum was also recorded with an air-filled cell, to measure the photoacoustic signal generated from the light absorbed by the cell windows and walls. In the \ce{NO2} spectrum shown in figure \ref{fig:monochrom}b, the air-filled cell spectrum has been subtracted to show the PA signal corresponding to \ce{NO2} alone. The power-normalized signal is shown in the inset of figure \ref{fig:monochrom}b. The signal-to-noise ratio was limited due to low power throughput with the narrow bandwidth. The spectral shape matches the \ce{NO2} absorption cross section spectrum \cite{BURROWS98,Keller13}, with no deviations due to photodissosiation over the wavelength region with appreciable LED power.

\begin{figure}[h!]
    \centering
    \includegraphics[width=0.95\linewidth]{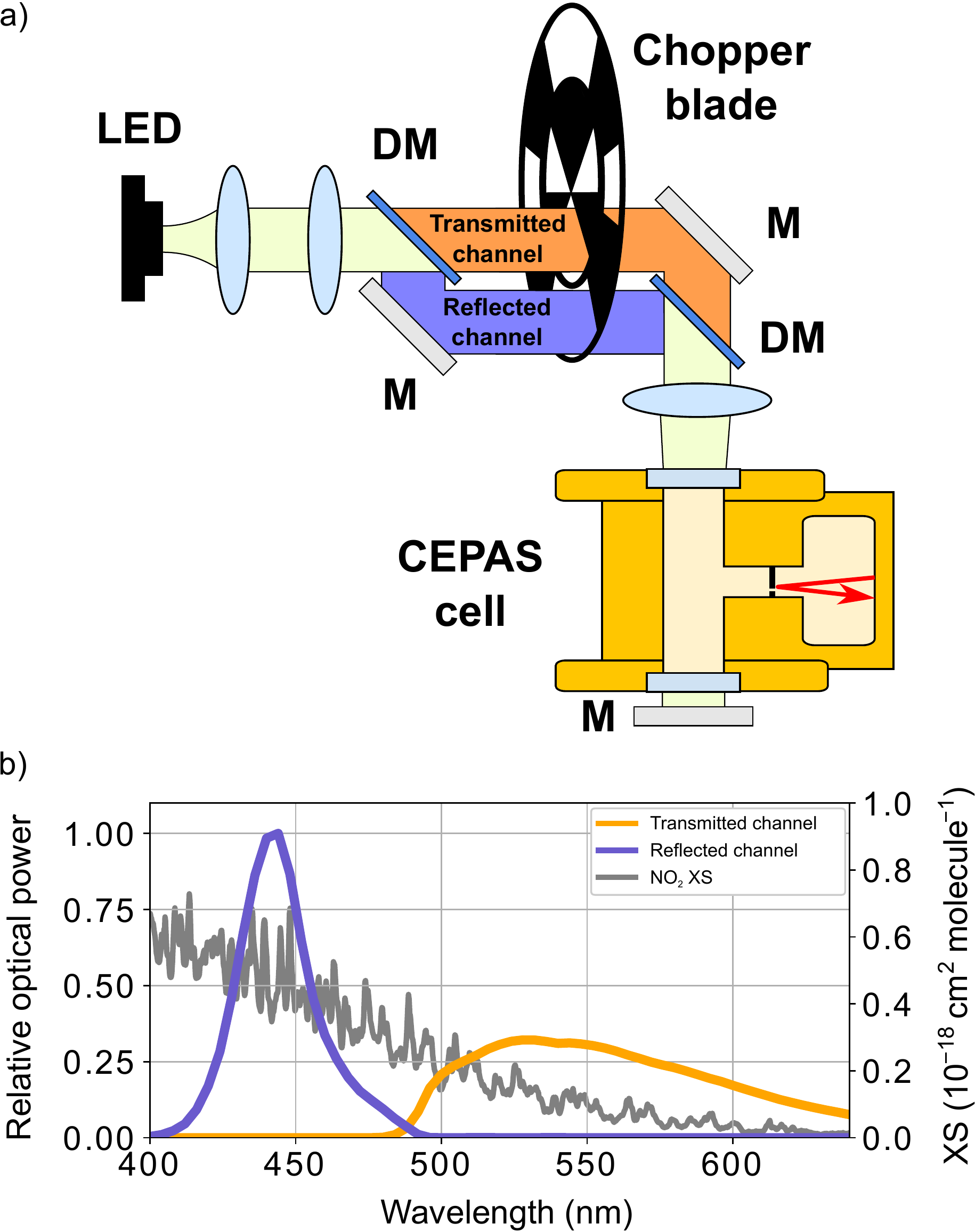}
    \caption{(a) Schematic picture of the two-channel \ce{NO2} concentration measurement setup. DM: Dichroic beamsplitter, M: aluminium mirror. The red arrow denotes the cantilever optical read-out system. (b) Spectrum of the relative LED optical power in each wavelength branch, calculated as a product of the LED spectrum measured with the monochromator (figure \ref{fig:monochrom}a) and the reflectivity or transmission of two dichroic mirrors as given by the manufacturer. The absorption cross-section of \ce{NO2} is also shown for reference \cite{BURROWS98}.}
    \label{fig:setup}
\end{figure}

After the verification of the spectral response, \ce{NO2} concentration measurements were performed with a two-channel configuration. A schematic picture of the concentration measurement setup is shown in figure \ref{fig:setup}a. The LED spectrum was divided into two parts with a dichroic beamsplitter (DLMP490R, Thorlabs). The light reflected by the beamsplitter is referred to as the reflected channel and the transmitted light is referred to as the transmitted channel. Figure \ref{fig:setup}b shows the spectrum of the relative LED power in each channel and how they relate to the \ce{NO2} absorption cross section. The separated beams were sent through different pitches of a dual-beam chopper blade to perform modulation frequency multiplexing. The channels were superimposed with another identical beamsplitter and focused into the CEPAS cell. The total optical power in most of the measurements, as measured before the cell input window and including the \SI{50}{\percent} duty cycle of the chopper, was approximately \SI{100}{\milli\watt}. Power measured after the cell output window was approximately \SI{83}{\percent} of the power measured before the input window. To increase the effective optical power coupled into the cell, a planar aluminium mirror was added after cell to reflect part of the transmitted light back into the cell. The chopper provides reference signals for both modulation frequencies of the dual-beam chopper blade, which in our measurements were \SI{267}{\hertz} and \SI{223}{\hertz}. The PA signal was demodulated at the corresponding frequencies. Sample gas inside the CEPAS cell was exchanged every 20 seconds. During a gas exchange, the cell was flushed for \SI{6}{\second} and left to stabilize for \SI{4}{\second}. After the gas exchange, the signals were recorded for \SI{10}{\second}. The concentration measurements were performed at atmospheric pressure.  Different \ce{NO2} concentrations used in calibration were mixed down with synthetic air from a gas cylinder containing \SI{10}{ppm} of \ce{NO2} in nitrogen. The mixing was done with mass flow controllers (SLA5850, Brooks Instrument).

The photoacoustic signal generated by the light reflected from the dichroic mirror follows the relation:\\
\begin{altequation}
    S_R=G_{A}G_{P}(f_{m})\int\left[\alpha(\lambda)c_{NO_2}+\beta(\lambda)\right]P(\lambda)R(\lambda) \text{d}\lambda .
\end{altequation}
Here, $\alpha$ is the absorption coefficient of \ce{NO2}, integrated over the length of the cell, $c_{NO_2}$ is the concentration of \ce{NO2}, $\beta$ is an absorption coefficient for the cell walls and windows, $P$ is the optical power emitted by the LED, $R$ is a factor describing the total throughput of the light reflected by the dichroic mirrors, $G_{A}$ is the gain of the data acquisition system, and $G_{P}$ is the responsivity of the photoacoustic cell, which depends on the modulation frequency $f_{m}$. We assume that for small changes in the LED power, the emission spectrum is constant, so that the total power $P$ can be taken out from the integral. That is, $P(\lambda)=P\times \phi_P(\lambda)$, where $\phi_P$ gives the emission wavelength distribution, with its integral over the wavelength normalized to unity. Later on we will consider the limitations caused by the simplifications done in the model. The signal is thus given by:
\begin{altequation}
    \begin{gathered}
        S_R=PG_{A}G_{P}(f_{m})\left(\int\left(\alpha(\lambda)\phi_P(\lambda)R(\lambda)\right)\text{d}\lambda \times c_{NO_2}\right.\\
        \left.+\int\left(\beta(\lambda)\phi_P(\lambda)R(\lambda)\right) \text{d}\lambda\right).
    \end{gathered}
\end{altequation}
Within these assumptions, the PA signal depends on the \ce{NO2} concentration according to the relation:
\begin{altequation}
    S_R=PG_{A}\left(\alpha_R\times c_{NO_2}+\beta_R\right).
\end{altequation}
The coefficients $\alpha_R$ and $\beta_R$ are the absorption coefficients integrated over the wavelength distribution of the light reflected by the dichroic mirror. Similarly, by replacing the throughput for reflectivity branch $R(\lambda)$ with that of the transmission branch $T(\lambda)$, we get the PA signal generated by light transmitted through the dichroic mirrors:\\
\begin{altequation}
    S_T=PG_{A}\left(\alpha_T\times c_{NO_2}+\beta_T\right).
\end{altequation}

The photoacoustic response has been incorporated within the coefficients $\alpha_R$, $\beta_R$, $\alpha_T$ and $\beta_T$, because it can be different for the two branches as it depends on the modulation frequency. Furthermore, the coefficients are affected by any difference in alignment and focusing between the branches, which may change the relative power absorbed by the gas phase and the cell walls. The quotient of the signals from the two branches can be used to deduce the \ce{NO2} concentration and it suppresses common mode noise and signal drifts originating from the LED power or the responsivity of the acquisition system:
\begin{altequation}
    \dfrac{S_R}{S_T}=\dfrac{\dfrac{\alpha_R}{\beta_T}\times c_{NO_2}+\dfrac{\beta_R}{\beta_T}}{\dfrac{\alpha_T}{\beta_T}\times c_{NO_2}+1}.
    \label{eq:model}
\end{altequation}
In the concentration measurements, the dichroic mirror divided the spectrum so that the reflection channel contained the narrow peak at \SI{440}{\nano\meter} and the transmission channel contained the wider peak which has its maximum at approximately \SI{520}{\nano\meter} (see figure \ref{fig:setup}b). The reflection channel thus had substantially stronger absorption by \ce{NO2}.

\section{Results and discussion}
\begin{figure}[b!]
    \centering
    \includegraphics[width=0.95\linewidth]{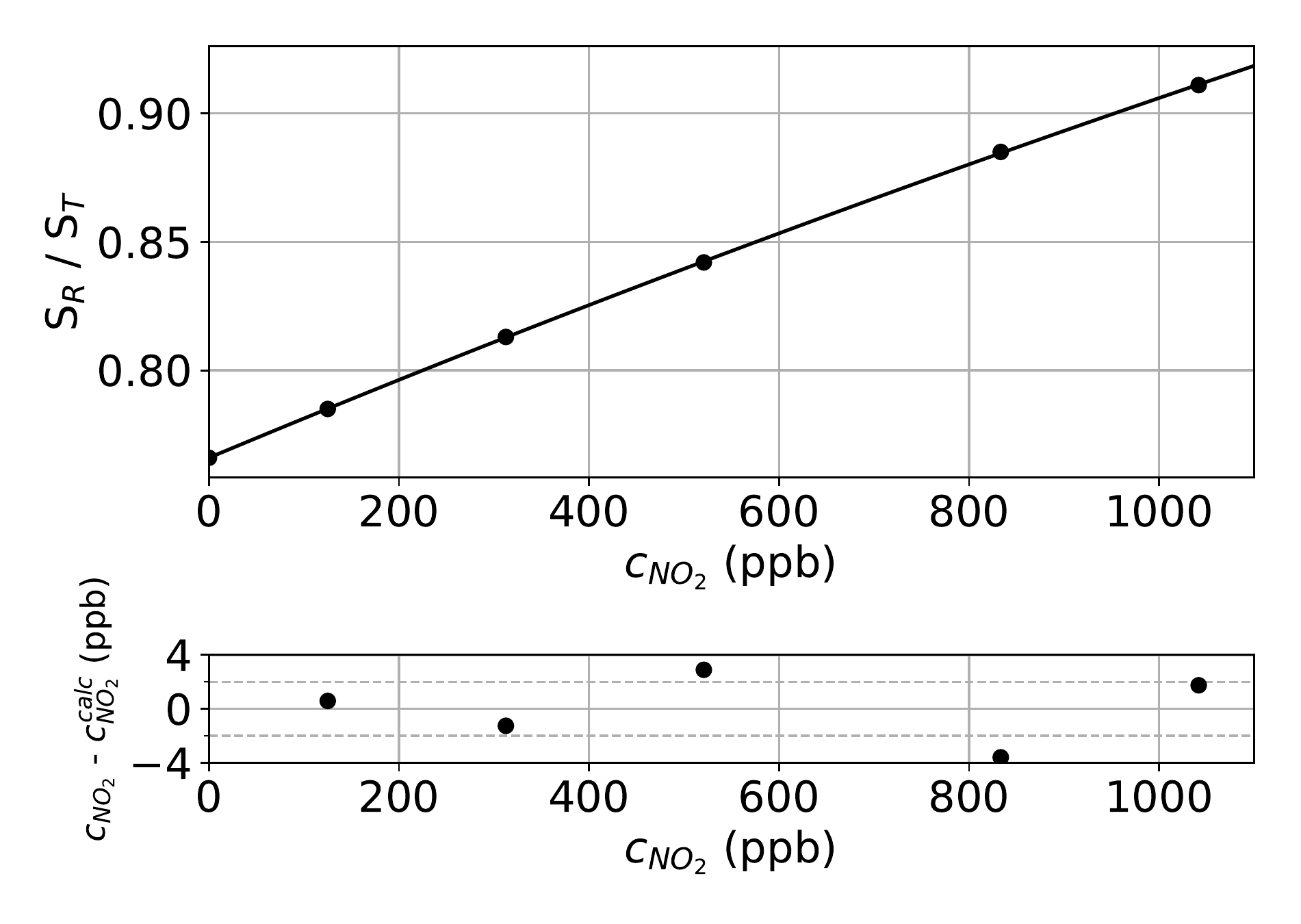}
    \caption{The photoacoustic signal for the calibration set, calculated as a quotient of the two channels, and a least-squares fit using equation \eqref{eq:model} (above). The difference between the concentration set with the mass flow controllers and the concentration calculated with the fitted model is also shown (below).}
    \label{fig:calibration}
\end{figure}

A calibration measurement set was performed to fix the coefficients in equation \eqref{eq:model}. The coefficient $\sfrac{\beta_R}{\beta_T}$ was fixed by recording the average signal from a long measurement of synthetic air without any \ce{NO2}. The other coefficients $\sfrac{\alpha_R}{\beta_T}$ and $\sfrac{\alpha_T}{\beta_T}$ were calculated with a least-squares fit to a set of \ce{NO2} concentrations ranging from \SI{125}{ppb} to \SI{1.042}{ppm}, using the Levenberg-Marquardt algorithm. The \ce{NO2} concentration was first set to \SI{2}{ppm} for \SI{20}{\minute} in order to minimize errors arising from \ce{NO2} lost to the gas line walls. For the subsequent lower concentration levels, which were used as the calibration set, the mixed gas flow was let to stabilize for \SI{5}{\minute} and then an average over \SI{10}{\minute} was recorded. The optical power passing into the CEPAS cell was approximately \SI{100}{\milli\watt}. Figure \ref{fig:calibration} shows the fit to the calibration data set. After the coefficients in equation \eqref{eq:model} were fixed, the \ce{NO2} concentration can be solved as a function of the quotient $\sfrac{S_R}{S_T}$. The differences between the concentrations set by the mass flow controllers and the concentrations calculated using the fitted coefficients are also shown in figure \ref{fig:calibration} for each calibration step.

\begin{figure}
    \centering
    \includegraphics[width=0.95\linewidth]{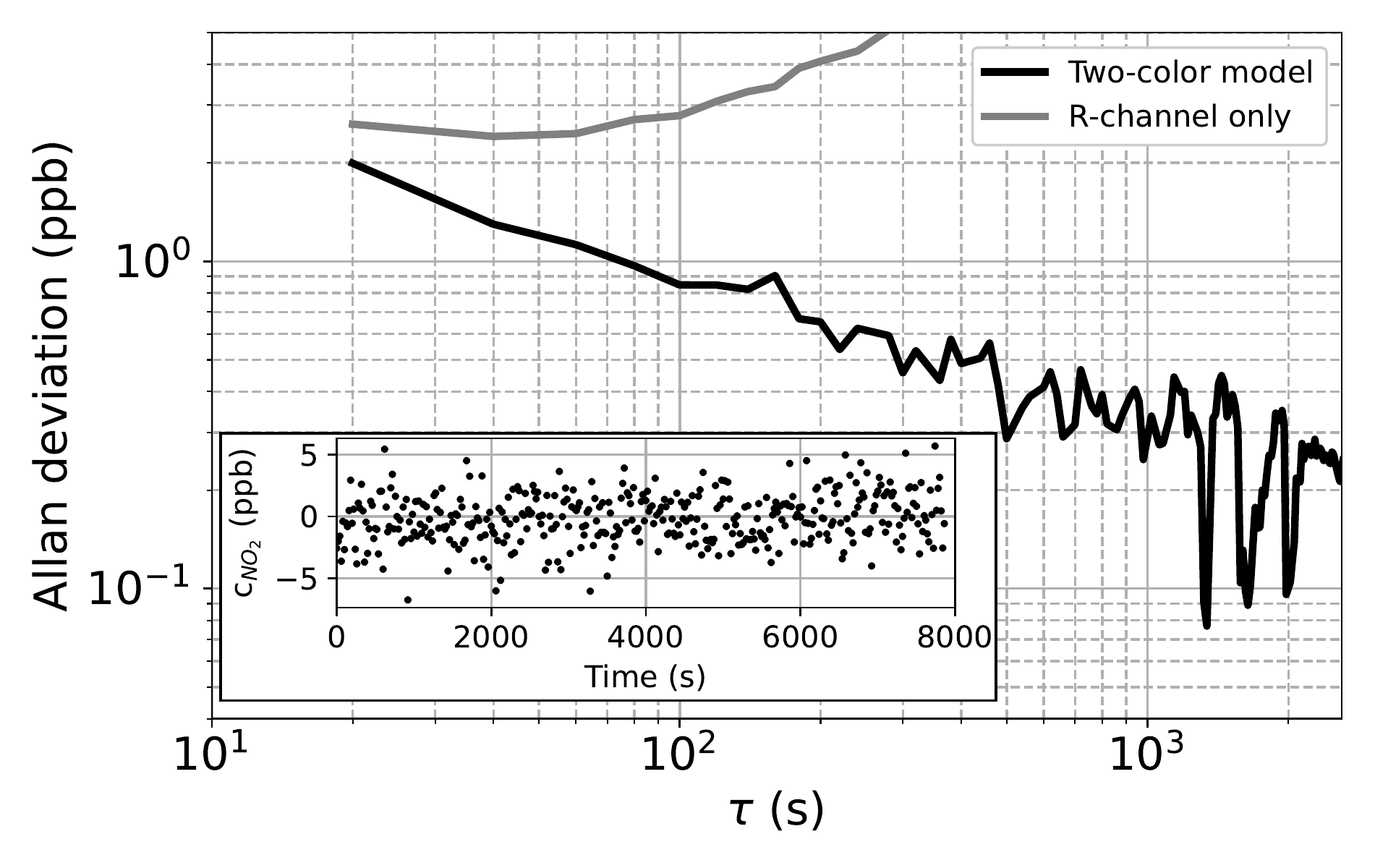}
    \caption{Allan-Werle deviation of concentration from a background measurement with synthetic air, calculated with the model given by equation \eqref{eq:model} (Two-color model). Allan-Werle deviation for concentration estimate calculated with $S_R$ only is also shown for reference  (R-channel only). The concentration estimate from $S_R$ is calculated based on the signal level during this synthetic air measurement and the signal level with \SI{1.042}{ppm} of \ce{NO2} in the calibration set. The inset shows the time trace for the concentration calculated with the two-color model.}
    \label{fig:allan}
\end{figure}

The long measurement that was used to fix the coefficient $\sfrac{\beta_R}{\beta_T}$ was also used to evaluate the stability of the measurement system. The setup was first let to stabilize for \SI{30}{\minute} and then the signal from synthetic air was recorded for \SI{2}{\hour}. Figure \ref{fig:allan} shows the concentration calculated with the model over time, as well as the Allan-Werle deviation of the time trace \cite{werle93}. A single data point is the average over the \SI{10}{\second} data acquisition time between each gas exchange. One measurement thus takes 20 seconds and figure \ref{fig:allan} shows that noise level between successive measurements corresponds to a \ce{NO2} concentration of \SI{2}{ppb}. A noise level ($1\sigma$) below \SI{1}{ppb} is reached after \SI{70}{\second}. No drifts were observed over the period of \SI{2}{\hour}. For reference, the standard deviation over the \SI{10}{\minute} measurement of the largest concentration in the calibration set (\SI{1.042}{ppm}) was \SI{2.6}{ppb}. This is similar to the noise level of the blank measurement as given by the Allan-Werle deviation and shows that the measurement precision is not impaired much by the presence of the analyte or the higher signal level.

In the model, the LED emission spectrum was assumed to be independent of the driving current. In practise, the spectrum can change since the LED emission wavelength is a function of its temperature and the efficiency of the phosphor may depend on temperature and excitation power. To test the effect that variations in the LED power has on the measured concentration, LED drive current was varied from zero to the maximum current rating of the LED. We observed a linear dependence between the LED optical power and the quotient signal, which translated to approximately \SI{6}{ppb/mW} at zero concentration. The stability of the LED power was tested over a measurement period of half an hour and the relative stability was better than $10^{-4}$ with the power level used in the concentration measurements. For accurate measurements, the detection system should be calibrated to the used emission power, but the LED power stability has only a vanishing effect on the concentration measurement.

The response of the CEPAS cell is a function of the modulation frequency and drifting of the oscillator driving the chopper may affect the concentration signal. The quotient signal reduces the effect over small changes because the two frequencies are derived from the same oscillator and change together, but it is not removed completely because the dependence is not linear. By varying the modulation frequency, we observed that around the used frequencies, a change of \SI{10}{\hertz} caused a change of \SI{5}{ppb} in the measured concentration. Observed drifts in the measured modulation frequency were up to the order of \SI{0.5}{\hertz}, which would produce an error of only \SI{0.25}{ppb}.

Interference from water was also evaluated. Water can have a complex interference to the concentration signal, since it can affect the acoustics of the photoacoustic cell as well as the relaxation rate of excited states in air \cite{RUCK17,YIN17}. To evaluate the effect of water concentration, the humidity of the sample gas was varied by passing part of the synthetic air flow over the air space above distilled water. By increasing the water concentration from 0 to \SI{1}{\percent}, the raw photoacoustic signals increased by approximately \SI{2}{\percent}. The increase was similar when \ce{NO2} concentration was 0 or \SI{100}{ppb}. However, the increase was not equal in both branches and the concentration calculated with the model decreased by approximately \SI{30}{ppb}. At least part of the decrease is caused by optical absorption, since the higher wavelength tail of the LED spectrum reaches relatively strong water lines. An estimation using lines listed in the HITRAN database \cite{GORDON17} predicts that the increased signal of the transmission branch due to \SI{1}{\percent} of water vapor decreases the calculated concentration by approximately \SI{10}{ppb} when the actual concentration is 0. Although the change from 0 to \SI{1}{\percent} in absolute humidity is drastic for ambient air, atmospheric \ce{NO2} measurements at ppb-level would benefit from either dehumidifying the measured samples, or a simultaneous measurement of the humidity and a correction to the photoacoustic signal. Interference from water absorption could also be reduced by filtering a narrower portion of the LED spectrum, but this would be done at a cost of sensitivity due to reduced optical throughput.

\section{Conclusions}
In conclusion, we have demonstrated a simple LED based photoacoustic \ce{NO2} sensor with a sub-ppb detection limit. The performance is high enough to detect \ce{NO2} concentrations, which are lower than its typical atmospheric abundance. The setup could alternatively use two different LEDs instead of the dichroic filtering. This could improve selectivity, but may reduce the cancellation of common drifts, since the emission power of two LEDs might not be as highly correlated. The measurement setup also demonstrates that with CEPAS, low detection limits can be reached with visible LED sources. The method could be extended to high-sensitivity detection of other chemical compounds, which have high absorbance at visible and ultraviolet wavelengths. As LEDs are available at wide selection of emission wavelengths, the method offers potential for development of cost-effective and sensitive detectors for a variety of trace gasses.
\section*{Acknowledgments}
The work was supported by the Academy of Finland (Project numbers 326444 and 314364) and by the Academy of Finland Flagship Programme, Photonics Research and Innovation (PREIN), decision number: 320167.

\bibliographystyle{ieeetr}
\bibliography{bibliography}
\end{document}